\documentclass[10pt,journal,compsoc]{IEEEtran}
\IEEEoverridecommandlockouts

\usepackage{pifont} 

\usepackage{cite}
\usepackage{amsmath,amssymb,amsfonts}
\usepackage{algorithmic}
\usepackage{graphicx}
\usepackage{textcomp}
\usepackage[dvipsnames]{xcolor}
\usepackage{array}
\usepackage{url}
\usepackage{multirow}
\usepackage{verbatim}

\usepackage[centerlast,small,sc]{caption}

\usepackage{hyperref}
\hypersetup{
    colorlinks=true,
    linkcolor=blue,
    filecolor=magenta,      
    urlcolor=BlueViolet,
    pdftitle={Sharelatex Example},
    pdfpagemode=FullScreen,
    }

\def\BibTeX{{\rm B\kern-.05em{\sc i\kern-.025em b}\kern-.08em
    T\kern-.1667em\lower.7ex\hbox{E}\kern-.125emX}}

\usepackage{tikz}
\newcommand*\circled[1]{\tikz[baseline=(char.base)]{\textbf{
            \node[shape=circle,draw,inner sep=1pt] (char) {#1};}}}

\newcommand{\ie}{\unskip, \textit{i.e.},\ }

\newcommand{\etal}{\textit{et al}. }

\newcommand{\quotes}[1]{``#1''}

\makeatletter
\def\thickhline{%
  \noalign{\ifnum0=`}\fi\hrule \@height \thickarrayrulewidth \futurelet
   \reserved@a\@xthickhline}
\def\@xthickhline{\ifx\reserved@a\thickhline
               \vskip\doublerulesep
               \vskip-\thickarrayrulewidth
             \fi
      \ifnum0=`{\fi}}
\makeatother
\newlength{\thickarrayrulewidth}
\makeatletter
\def\thickerhline{%
  \noalign{\ifnum0=`}\fi\hrule \@height \thickerarrayrulewidth \futurelet
   \reserved@a\@xthickerhline}
\def\@xthickerhline{\ifx\reserved@a\thickerhline
               \vskip\doublerulesep
               \vskip-\thickerarrayrulewidth
             \fi
      \ifnum0=`{\fi}}
\makeatother
\newlength{\thickerarrayrulewidth}
\begin{document}

\title{A Novel Approach to Identify Security Controls in Source Code}

\author{Ahmet~Okutan,
Ali~Shokri,
Viktoria~Koscinski,
Mohamad~Fazelinia,
Mehdi~Mirakhorli
\IEEEcompsocitemizethanks{\IEEEcompsocthanksitem All authors are with the Department of Software Engineering, Rochester Institute of Technology, Rochester,
NY, 14620. (email: ahmet.okutan@gmail.com; as8308@rit.edu; vk2635@rit.edu; mf8754@rit.edu; mxmvse@rit.edu)}
\thanks{}}


\IEEEtitleabstractindextext{

\begin{abstract}
\textit{Secure by Design} has  become  the  mainstream  development  approach ensuring that software systems are not vulnerable to cyberattacks. Architectural security controls need to be carefully monitored over the software development life cycle to avoid critical design flaws. Unfortunately, functional requirements usually get in the way of the security features, and the development team may not correctly address critical security requirements. Identifying tactic-related code pieces in a software project enables an efficient review of the security controls' implementation as well as a resilient software architecture. This paper enumerates a comprehensive list of commonly used security controls and creates a dataset for each one of them by pulling related and unrelated code snippets from the open API of the {\texttt{StackOverflow}} question and answer platform. It uses the state-of-the-art NLP technique Bidirectional Encoder Representations from Transformers (BERT) and the Tactic Detector from our prior work to show that code pieces that implement security controls could be identified with high confidence. The results show that our model trained on tactic-related and unrelated code snippets derived from \texttt{StackOverflow} is able to identify tactic-related code pieces with F-Measure values above 0.9.   
\end{abstract}

}

\maketitle

\section{Introduction}
\textit{Secure by Design} has become the mainstream development approach to ensure the security and privacy of software systems. During the software development life cycle, security requirements are identified early and addressed from the ground up with a robust architectural design. A secure design often relies on well-known \textit{security controls}\cite{Bass}, defined as reusable techniques to achieve specific quality concerns. These controls provide solutions to enforce \textit{authentication}, \textit{authorization}, \textit{confidentiality}, \textit{data integrity}, \textit{privacy}, \textit{accountability}, \textit{availability}, \textit{safety} and \textit{non-repudiation} requirements, even when the system is under attack \cite{Hafiz:2012}.

Architectural security controls need to be identified in the early stages of the design process and then implemented alongside functional features \cite{ATAM, TaylorBook, Fairbanks, Perry1992}. However, in practice, this is not always the case, as architectural solutions often evolve during the development process \cite{Bosch, TwinPeaks}. For example, the analysis of the source code of the \emph{Wavelet Enterprise Management Portal (EMP)} \footnote{https://wavelet.net/erp-emp-training/}, a large open-source enterprise resource planning and customer relationship management system, shows that its initial release included a module to edit sensitive personal information including social security number and credit limit. However, no \emph{Audit Trail} control was provided until the system was refactored in the version 6 release. This happened due to developers' primary focus on implementing the functional requirements, which lead to ignoring architecturally significant security requirements either intentionally or accidentally and not incorporating appropriate security controls to satisfy quality concerns in a timely manner. 

In practice, many software projects fail to fully document security-related architectural design decisions, providing only high-level lists of decisions, if any. Existing approaches generally document rationales at the design level, and fail to provide explicit traceability down to the code level in which the security controls have been implemented \cite{Mirakhorli-Cleland-2011}. Therefore, the state of the practice solutions provide only limited support for keeping developers informed of underlying security design decisions during the code maintenance process.

To address this problem, we present an automated and \textit{just-in-time} reverse engineering solution to detect architectural security controls implemented in the Java language. This solution identifies security controls incorporated by designers and pinpoints modules and source files that implement them. This will help developers to understand  underlying security design decisions as they sustain a software product~\cite{Mirakhorli2016}. We use the Bidirectional Encoder Representations from Transformers (BERT) \cite{bert-2018} and the Tactic Detector from our earlier work \cite{Mirakhorli2016} to identify security control-related code snippets and map them to a security control catalog to help developers have a better understanding of the security design decisions implemented in their code.

\begin{figure}[!h]
\centering
\includegraphics[width=\columnwidth]{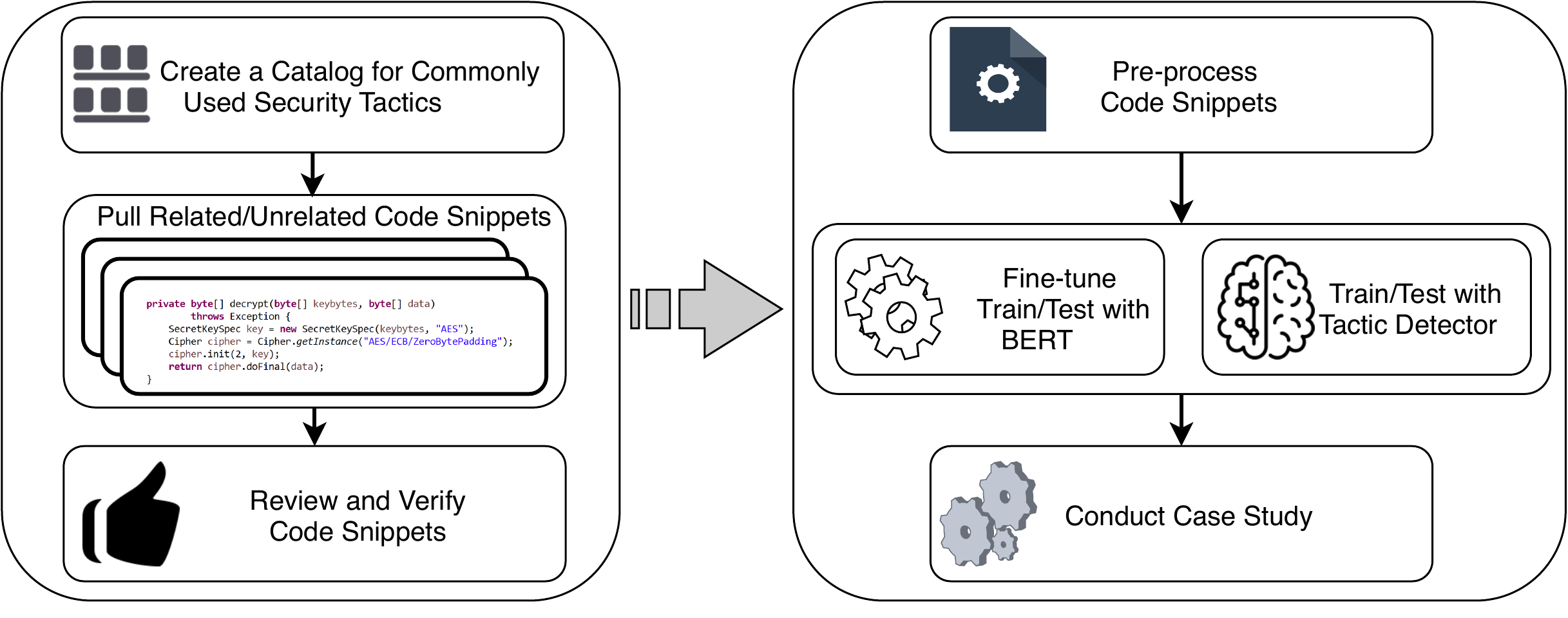}
\caption{An overview of the tactic selection, data set generation, and tactic classification processes used in the paper.}
\label{fig:overall_summary}
\end{figure}

Experiment results indicate that code snippets that implement security tactics can be identified with F-Measure values up to 0.98. The main contributions of this paper are:
\begin{itemize}
    \item This paper defines three high-level security control categories \ie \textit{Detect}, \textit{Prevent}, and \textit{React} and compiles a comprehensive list of commonly used security controls to fill each category.
    
    \item It presents and evaluates a novel methodology and algorithms to identify security control-related code snippets in a given software system. A state-of-the-art classification method is used to identify code snippets related to security controls with high Precision, Recall, and F-Measure values. To the best of our knowledge, no \textit{transformer models} were used to classify security tactics before this study. 
    
    \item Finally, we provide an online appendix that includes our security control catalog, a \textit{replication package} containing our dataset and scripts used to fine-tune, train and test our models, evaluation pipelines, and a supplementary reference data: 
    \textbf{\url{https://github.com/SoftwareDesignLab/security_tactics}}

\end{itemize}

The rest of the paper is organized as follows: Section \ref{related_work} reviews prior work about identifying design patterns and architectural tactics from the source code and provides a brief overview of the evolution of NLP methods and the BERT algorithm. Section \ref{methodology} explains the approach used to select commonly used security tactics and review derived code snippets for each security tactic. Section \ref{experiments_results} presents the experiment results with a discussion of potential practical use cases. Section \ref{Threats_to_Validity} discusses threats to the validity of this work, and finally, Section \ref{Conclusion} provides the concluding remarks.

\section{Prior Work}
\label{related_work}
\begin{figure*}[h!]
\centering
\includegraphics[width=\textwidth]{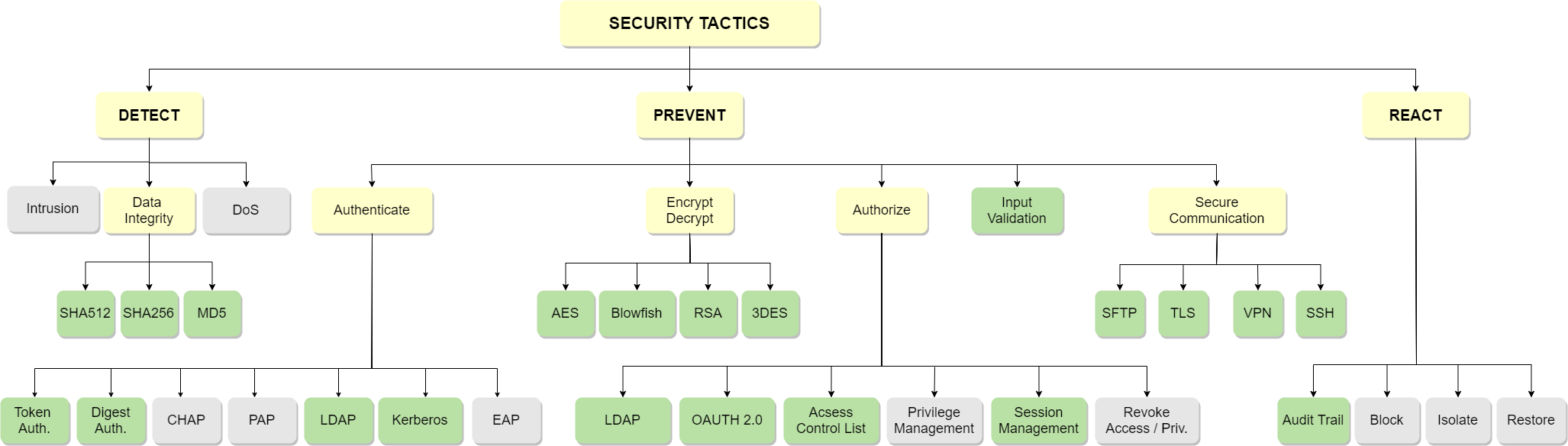}
\caption{Categorization of studied security controls. Yellow nodes show tactic categories. Green nodes represent security controls that have at least 25 related code snippets in Stack Overflow. Gray nodes represent security controls that had less than 25 related code snippets and are not included in experiments.}
\label{fig:tactic_categorization_chart}
\end{figure*}

Design patterns provide reusable solutions for commonly encountered problems during the system and object design activities. The use of design patterns helps to enhance the quality of the source code, thereby improve its resilience to architectural flaws. Detecting design patterns\cite{gamma-et-al} implemented in source code is useful for applications like reverse engineering, software maintenance, system comprehension, and documentation\cite{survey-2016,ZANONI2015102}. Over the years, researchers have developed different approaches to detect design patterns in the source code. These approaches mainly consist of two phases \ie the feature extraction and the design pattern detection. In the feature extraction phase, high-level abstract features are extracted from the source code or its intermediate representation, where intermediate representations could be an Abstract Syntax Tree (AST)\cite{Fontana2011ATF, ZANONI2015102}, Abstract Semantic Graph (ASG)\cite{1235436} or a set of source code metrics \cite{Dwivedi2016SoftwareDP}. Extracted features capture essential attributes of the underlying design patterns and are used in the pattern detection phase to identify whether a design pattern is implemented or not. Leveraging the most recent state-of-the-art architectures of NLP, this work generates context-aware vector representations from the code snippets and then uses them to train Machine Learning models to identify tactic related code pieces.

Architectural security controls play a key role in addressing reliability, performance, and security-related issues in software systems. Prior works have shown that the discovery of architectural security controls from the source code is possible using a variety of Machine Learning methods. Mirakhorli and Cleland-Huang \cite{Mirakhorli2016} use a customized classifier trained on code snippets extracted from fifty open source software systems and show that commonly referred architectural security controls representing security, performance, and reliability aspects can be identified with recall rates above 70 percent. Another study \cite{Mirakhorli2012} uses Machine Learning methods and lightweight structural analysis on 10 open source projects and achieves F-Measure values ranging between 0.59 and 0.96 while identifying heartbeat, scheduling, resource pooling, authentication, and audit trail related code snippets. Drawing inspiration from previous works in the field, this paper focuses specifically on the security domain, and trains the state-of-the-art text classification method BERT on the code snippets derived from Stack Overflow to show that security-related code snippets can be identified with relatively high accuracies.

Traditional token-based (bag-of-words) methods that have had difficulty in differentiating text pieces with different ordering of tokens have evolved to superior language models that use self-attention to learn better language representations \cite{vaswani2017attention}. Similar to the use of word embeddings on regular text\cite{mikolov2013efficient}, there is an increasing interest in learning code embeddings using vector representations of code-snippets\cite{chen2019literature}. Word embeddings have revolutionized NLP, leading to the development of pre-trained language models\cite{bert-2018, liu2019roberta, yang2019xlnet} that have achieved state of the art results in many NLP tasks. Similarly, effective techniques have been developed to learn code embeddings. These embeddings have been shown to capture semantic properties of the code-snippets and provide promising results for method and class name prediction\cite{alon2019code2vec, allamanis_et_al}, semantic code search\cite{10.1145/3211346.3211353, dcs, husain2019codesearchnet}, code generation\cite{ling2016latent}, code summarization\cite{iyer-etal-2016-summarizing}, and several other tasks\cite{survey_big_code}. Building upon the evolutionary improvements in the NLP domain so far, this work proposes to use Bidirectional Encoder Representations from Transformers (BERT) to learn source code representations from code snippets, and then, use the learned representations as features to detect whether a security tactic is implemented or not.

\subsection{Bidirectional Encoder Representations from Transformers (BERT)}
\label{bert}
Language models are trained on large data sets to learn word embeddings that can be used for various tasks. Word embeddings were popularized by Word2vec \cite{word2vec} and Global Vectors for Word Representation (Glove)\cite{glove}. These unidirectional models were trained on a large unlabelled corpus to learn representations that capture the semantic similarities between words. One limitation with these approaches is that they learn a single representation for each word in the corpus irrespective of the context in which the word is used. Embeddings from Language Models (ELMo)\cite{elmo} tries to solve this problem by generating contextual embeddings. ELMo trains a deep bidirectional LSTM on the language modeling task to embed the surrounding context of a word in its embedding. Then, the pre-trained ELMo language model can be used in other models that process text. This approach has achieved state-of-the-art results in many NLP tasks. However, the use of pre-trained language models is limited to the learned word embeddings. These word embeddings are used as fixed-parameters and the model needs to be trained from scratch for each task-specific dataset. There was no effective transfer learning technique to fine-tune these models on a different task. Universal Language Model Fine-Tuning (ULMFiT)\cite{howard2018universal} solves this problem by introducing a language model fine-tuning strategy to perform transfer learning. OpenAI's GPT-2\cite{gpt-2} replaces the LSTM blocks in language models with the decoder blocks from the Transformer model\cite{vaswani2017attention} and enables the language model to take advantage of the self-attention mechanism to learn better representations.
 
BERT\cite{bert-2018}, a pre-trained language model that can be fine-tuned to learn tasks with a fair amount of data, builds upon the recent developments in natural language processing. A transformer encoder\cite{vaswani2017attention} is bi-directionally trained using {\texttt{Wikipedia}} (\href{https://wikipedia.com}{https://wikipedia.com}) and the {\texttt{Book Corpora}} (\href{https://yknzhu.wixsite.com/mbweb}{https://yknzhu.wixsite.com/mbweb}).

BERT is pre-trained on two tasks:
\begin{itemize}
\item\emph{Masked Language Modeling}: 
15\% of input tokens are randomly replaced by the [MASK] token and then the model attempts to predict the original token from the final output embeddings corresponding to those masked tokens. This enables the model to learn relationships between words within a sentence.

\item\emph{Next Sentence Prediction}: 
In next sentence prediction, the model is given two sentences $A$ and $B$ and it attempts to predict whether sentence $B$ is the subsequent sentence in the corpus. The model uses the final output embedding corresponding to the first input token [CLS] for next sentence prediction. This enables the model to learn relationships between two sentences.
\end{itemize}

BERT has achieved state-of-the-art results on eleven NLP tasks\cite{wang-etal-2018-glue}. It has been shown that pre-training enables the model to learn better contextual embeddings. To use BERT for downstream tasks, it is fine-tuned on a new labelled task specific dataset and then used to solve problems like question-answering and classification. 

Larson \etal use BERT and a set of additional classifiers including Support Vector Machine (SVM), Multi Layer Perceptron (MLP), and Convolutional Neural Network (CNN) to analyze intent classification and out-of-scope prediction methods and state that BERT achieves the best in-scope accuracy, scoring 96\% or above even when the training data is limited \cite{Larson_bert_2019}. Another study \cite{ayse2020_bert} used part of the data set introduced by Larson \etal to compare the performance of BERT with simple LSTM models where the training data set included 15,101 entries for 150 intents (around 100 instances for each class). The results suggest that when less than 50\% of the training data is used (around less than 75 training instances for label), the performance of the BERT classifier gets worse, compared to a simple LSTM architecture with one bidirectional layer and one unidirectional layer with 50 neurons in each layer.

\section{Methodology}
\label{methodology}
This paper studies a list of security controls that are widely used by the software community. Inspired by the NIST Cybersecurity framework \cite{the_five_functions}, we define a high-level abstraction and categorize security controls into different groups according to the functions they are used for. Section \ref{tactic_selection} explains how security controls are selected and categorized, Section \ref{tactic_dataset} summarizes how the data set for each tactic is generated and manually reviewed. Finally, Section \ref{tactic_classification} explains how the code snippets are pre-processed, BERT models are fine-tuned, longer code snippets are treated with mean pooling, and BERT's output sequences are classified. Figures \ref{fig:dataset_generation} and \ref{fig:visual_overview} provide a high-level overview of these processes explained in Sections \ref{tactic_selection}, \ref{tactic_dataset}, and \ref{tactic_classification}.

\subsection{Tactic Selection}
\label{tactic_selection}
Security controls are selected to cover different functions defined by the NIST Cybersecurity framework about detecting, preventing, and reacting to a cybersecurity incident \cite{the_five_functions}. The controls are grouped into different categories based on their purpose of use. The categorization tree in Figure \ref{fig:tactic_categorization_chart} shows all studied categories and controls. A short description of each category and tactic is provided below:

\begin{figure}[!h]
\centering
\includegraphics[width=\columnwidth]{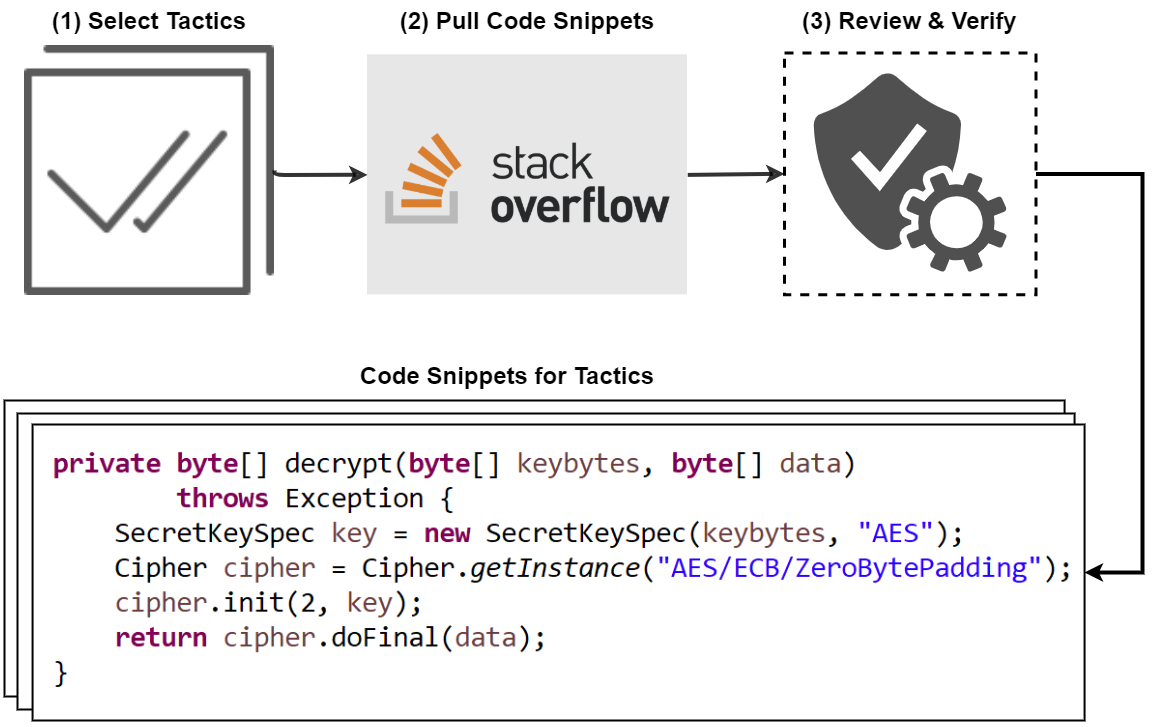}
\caption{An overview of the data set generation process.}
\label{fig:dataset_generation}
\end{figure}

\subsubsection{Detect}
\textit{Detect} is an abstract category that includes controls used to detect if a system or data has been compromised. The main focus is to detect anomalous cybersecurity events and make sure that protective measures are in effect. Security controls in this category are divided into three subcategories: 
\begin{itemize}
\item \textbf{Intrusion:} Intrusion controls monitor intrusion attempts, log information about them, and warn system administrators to take appropriate protective measures to address security issues.

\item \textbf{Data Integrity:} Data Integrity ensures that transmitted data is reliable and consistent throughout its life-cycle. A well-known approach to achieve data integrity is through Check-summing\cite{checksumming2015, checksumming}. A checksum of the data is calculated before and after the data is transmitted and the two check sums are compared to ensure that the data has not been modified. The most commonly used secure hash and checksum algorithms are SHA256, SHA512 \cite{sha_algs_Handschuh2005}, and MD5 \cite{md_algs_Bosselaers2005}.  

\item \textbf{Denial Of Service (DoS):}
This tactic monitors the system activities for Denial of Service attacks. Denial of service attacks flood the system with traffic and make it inaccessible to legitimate users\cite{dos_us_cert}. Denial of service attacks are handled by dropping illegitimate traffic and making the system available for legitimate users\cite{dos}.
\end{itemize}

\subsubsection{Prevent}
\textit{Prevent} category is defined to include controls used to ensure that a system is protected from cybersecurity incidents. This category is further divided into five different sub-categories, each representing a different aspect of the protection:

\begin{itemize}
\item \textbf{Authenticate:} Security controls in this category validate the identity of a user or a system. Commonly used examples include controls used for user authentication (Token Authentication and Digest Authentication), and network authentication (Kerberos and Light Weight Directory Access Protocol (LDAP)). Password Authentication Protocol (PAP), Challenge-Handshake Authentication Protocol (CHAP), and Extensible Authentication Protocol (EAP) are other methods that are used for authentication. 

\item \textbf{Encrypt/Decrypt:} Security controls in this group are used to secure data by transforming it into a format that cannot be read by eavesdroppers. Widely used encryption methods include Advanced Encryption Standard (AES), Blowfish, Rivest-Shamir-Adleman (RSA) and Triple Data Encryption Algorithm (3DES).

\item \textbf{Authorize:} This category includes controls used to validate access to system resources. Commonly used authorization controls include LDAP, OAuth 2.0, Access Control List, Privilege Management, Session Management, and Revoke Access/Privilege.

\item \textbf{Input Validation:} Input Validation is used to make sure that only properly formed and sanitized secure input is allowed to enter into a system.

\item \textbf{Secure Communication:} We define a new category named 'Secure Communication` to represent security controls that are used to establish end to end secure connections. Commonly used controls in this category are SSH File Transfer Protocol (SFTP), Transport Layer Security (TLS), Virtual Private Network (VPN) and Secure Shell (SSH).
\end{itemize}

\subsubsection{React}
This branch includes security controls used to respond or react to a cybersecurity incident. Once a cybersecurity event is detected, it is necessary to confine the impact of the event and restore the system capabilities that have been affected as soon as possible. An ideal cyber incident response should minimize the effects of the attack and find and fix its root cause to prevent future incidents. This tactic category is further divided into four sub-categories each representing a different task:
\begin{itemize}
\item \textbf{Audit Trail:} Audit Trail keeps records of the sequence of system events and user activities and these records can be used to identify malicious users, analyze security problems and detect violations. 

\item \textbf{Block: }  Cyber incidents can be contained by blocking known malware sources \ie domains or email addresses, closing related ports, tuning firewall filters to restrict access to certain software services that might be more vulnerable.  

\item \textbf{Isolate: } It is part of the containment process that makes sure that the effects of a cyber incident are confined, and damaged areas are isolated, with the incident investigated and the root causes identified.

\item \textbf{Restore: }Restore controls ensure that software systems or assets that are affected from a cybersecurity incident are adequately restored after the incident.  
\end{itemize}

\subsection{Dataset Generation}
\label{tactic_dataset}
This section explains how the code snippets for each security tactic are queried from Stack Overflow, how pulled code snippets are reviewed, and what type of security functions are included in each tactic data. An overview of these processes is shown in Figure \ref{fig:dataset_generation}. 

The data set for each security tactic is generated by extracting tactic related and unrelated code snippets from Stack Overflow. The advanced search API (\url{https://api.stackexchange.com/docs/advanced-search}) takes a free form text parameter and returns a list of questions. Stack Overflow matches the parameter with all question properties and returns the most relevant questions and answers. The questions and the topics tagged to them are curated and verified by the community, therefore we assume that the questions and the topics that they are tagged with are relevant. The advanced search API's underlying algorithm is undocumented; however, returned results show that it does not perform an exact string match only. For instance, we use "3DES" as the search parameter to search for questions related to the Triple DES tactic. However, returned questions include discussions that talk about 3DES, Triple DES and DESede. To get the maximum number of questions, we test multiple queries related to each tactic and choose the query that gives the highest number of relevant results.
For example, for tactic Advanced Encryption Standard (AES) we tried `advanced encryption standard' and `aes' and found that `aes' gives the most relevant results. The keywords used to query each security tactic are listed in Table \ref{tab:data_info}. 

As shown in Figure \ref{fig:dataset_generation}, after all of the questions are pulled for each security tactic, they are manually reviewed to ensure that the code-snippets associated to the questions are related to the queried tactic. To make sure that we have enough training samples for the training phase, we ignore security tactics that have lower than 25 related code snippets. Prior research\cite{Chaiyong2018} shows that a set of issues are observed in Stack Overflow answers including mismatched solutions, outdated solutions, incorrect solutions, and buggy code. Our finding, which is in line with this work, shows that although a sufficient number of questions are found for a security tactic and all pulled questions are associated with a variety of topics related to the tactic, the code-snippets included in the questions may not be related at all. Therefore, such code snippets are filtered out from the tactic related code snippets manually during the data review and verification process (\ie step 3 of Figure \ref{fig:dataset_generation}). In general, pulled code-snippets implement a wide variety of functions related to a security tactic. For instance, LDAP includes code pieces that implement LDAP authentication or query LDAP server for user or directory information. Similarly, Kerberos code snippets provide implementations of Kerberos authentication with three different frameworks. Below we provide a summary of the tasks (functions) implemented by the code-snippets included for each security tactic.

\subsubsection{Detect}
Under the \textit{Detect} category, we did not find enough samples for the \textbf{Intrusion and Denial of Service (DoS) controls.} For SHA-256, SHA-512 and MD5 message-digest controls, pulled code snippets include functions to hash data with MessageDigest (in the java.security package) or generate Hash-based Message Authentication Code (HMAC) SHA256 signature using the Java Cryptography Architecture (JCA) framework\cite{jca}.

\begin{table*}[h!]
\centering
\footnotesize
\begin{tabular}{ p{5cm}  p{3cm}  p{1cm}  p{1cm}  p{1cm}  p{1cm}  p{1cm}  p{1cm}}
 \thickerhline
  \textbf{Tactic}& \textbf{Query Parameter}& \textbf{Sample Size}&             \textbf{S1}& \textbf{S2}& \textbf{S3}& \textbf{S4}& \textbf{SMax}\\
  \thickhline
  SHA512 & sha512 & 100 & 98\%&	100\%&	-&	-&	2\\
  SHA256 & sha256 & 250 & 98.80\%&	100\%&	-&	-&	2\\
  MD5 & md5 & 400 & 98\%&	99.50\%&	100\%&	-&	3\\
  Token Authentication & token authentication & 200 & 98\%&	100\%&	-&	-&	2\\
  Digest Authentication & digest authentication & 100 &96\%&	99\%&	100\%&	-&	3\\
  LDAP & ldap & 400 & 99\%	&100\%	&-&	-&	2\\
  Kerberos & kerberos & 150 & 98\%	&100\%	&-&	-&	2\\
  AES & aes & 400 & 98\% & 100\% & - & - &2\\
  Blowfish & blowfish & 100 & 100\%	&-	&- &	-&	1\\
  RSA & rsa & 400  & 96.5\%&	100\%&	-&	-&	2\\
  3DES & 3des & 100 & 98\% & 100\% & - & - & 2\\
  OAuth 2.0 & oauth2 & 400 & 94.5\%&	99.75\%&	100\%&	-&	3\\
  Access Control List & acl & 150 & 100\% & - & - & - & 1\\
  Session Management & session management & 350  &97.71\%&	99.71\%&	99.71\%&	100\%&	4\\
  Input validation & validation interceptor & 50  & 98\%	&100\%	&-&	-&	1\\
  SFTP & sftp & 400 &97.75\%&	99.75\%&	100\%&	-&	3\\
  TLS & tls & 400 & 96.5\%&	99.75\%&	100\%&	-&	3\\
  VPN & vpn & 50   & 96\%&	100\%&	-&	-&	2\\
  SSH & ssh & 400 & 99.5\%&	99.5\%&	99.5\%&	100\%&	4\\
  Audit trail & audit trail & 200 & 98.75\%	&100\%&	-&	-	&2\\
  \thickerhline
\end{tabular}
    \caption{The summary of the tactic data sets used. S1, S2, S3, and S4 columns represent the percentage of code snippets that are divided into 1, 2, 3, and 4 sequences for BERT, respectively. The SMax column represents the number of sequences that the longest input is divided into.}
    \label{tab:data_info}
\end{table*}

\begin{figure*}[h!]
\centering
\includegraphics[width=\textwidth]{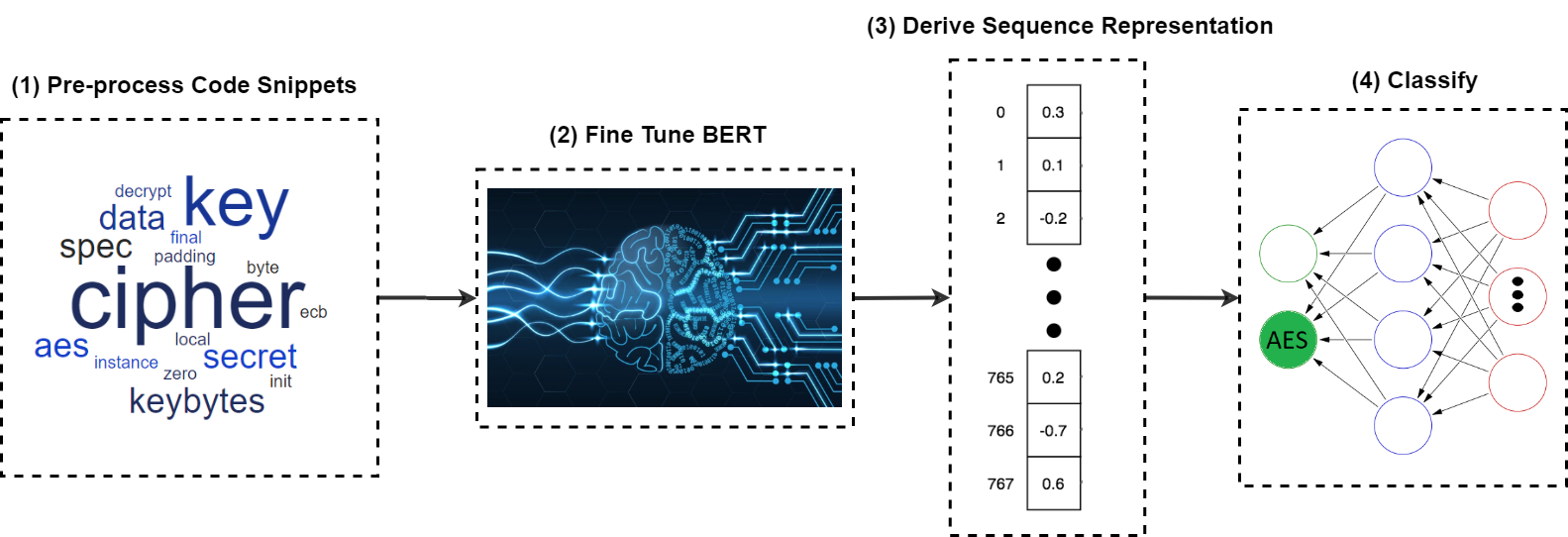}
\caption{Pre-process code snippets, fine-tune BERT and classify sequence representations.}
\label{fig:visual_overview}
\end{figure*}

\subsubsection{Prevent}
There are not enough samples for CHAP, PAP and EAP controls under `Authenticate' category. A summary of tasks included for other controls in this category are:
\begin{itemize}
    \item \textbf{Digest Authentication:} Code pieces that implement Spring Security Digest Authentication\cite{spring_security_digest_authentication} and Digest-MD5 authentication\cite{digest_md5_authentication}.
    
    \item \textbf{Kerberos:} Code snippets that implement Kerberos authentication using an extension of the Spring Security named Kerberos Spring Security\cite{kerberos_spring_security}, Java Authentication and Authorization Service (JAAS) Krb5LoginModule\cite{krb5loginmodule}, and the Generic Security Service API\cite{kerberos_gss_api}. Some of the code snippets include functions that use Kerberos to connect to Hadoop File System (HDFS).
    
    \item \textbf{Lightweight Directory Access Protocol (LDAP):} Functions that perform LDAP authentication and query LDAP server for user and directory information.
    
    \item \textbf{Token based authentication:} Code snippets that implement JSON Web Token (JWT)\cite{jwt} authentication using the Spring Security framework.
\end{itemize}

For the `Encrypt/Decrypt' category, a sufficient number of code snippets was found for each of the four controls. The functions included for each of these four controls are:  
\begin{itemize}
    \item \textbf{Advanced Encryption Standard (AES):} Encrypt or decrypt a password, message or file content using the Java Cryptographic Architecture (JCA) framework\cite{jca}.
    
    \item \textbf{Blowfish:} Encrypt or decrypt password, message or file content using the JCA framework.
    
    \item \textbf{Rivest-Shamir-Adleman (RSA):} Generate keys using the Java KeyPair class, and encrypt or decrypt a password, message or file content using the JCA framework.
    
    \item \textbf{Triple DES:} Encrypt or decrypt a password, message or file content using the JCA framework.
\end{itemize}

Under the `Authorize' category, there was not a sufficient number of samples for `Session Management' and `Revoke Access/Privilege' controls. The functions in the other four controls are:

\begin{itemize}
    \item \textbf{LDAP:} Perform LDAP authentication and query LDAP server for user and directory information.
    
    \item \textbf{OAuth 2.0:} Implement Oauth 2.0 using the Spring Security OAuth project\cite{spring_security_oauth}, access server resources with an access token.
    
    \item \textbf{Access Control List (ACL):} Implement access control using Spring Security ACL services\cite{spring_security} or Apache ZooKeeper\cite{apache_zookeper}.
    
    \item \textbf{Session Management:} Manage user session with Spring Session\cite{spring_session}, manage database session with Hibernate Session.
\end{itemize}

For `Input Validation', we include code snippets that validate arguments with Hibernate Validator and network requests with Spring Web Services Interceptor. For the `Secure Communication' category, the functions included in the code snippets of each tactic are:  
\begin{itemize}
    \item\textbf{SSH File Transfer Protocol (SFTP):} Access remote directory using Java Secure Channel\cite{java_secure_channel}, download remote files and upload files to a server.
    
    \item\textbf{Transport Layer Security (TLS):} Establish a TLS connection between client and server and perform certificate authentication.
    
    \item\textbf{Virtual Private Network (VPN):} Build VPN client using Android VPN Service and monitor network data.
    
    \item\textbf{Secure Shell (SSH):} Establish SSH connections using the Java Secure Channel framework and Java SSH client.
\end{itemize}

\subsubsection{React}
There are not enough code snippets for Block, Isolate and Restore controls in this category. The tasks included for the `Audit Trail' tactic are used to implement logging, transaction management, audit via the Hibernate Envers module\cite{hibernate_envers} and the Object Auditing and Diff Framework for Java (JaVers)\cite{javers}.

In addition to the tactic related code snippets, we include code snippets that are not related to the security controls. Unrelated code-snippets are extracted from a set of 50,000 questions with 'Java' tag. We use the generic keyword ‘Java’ to make sure that the questions are general and not biased towards any topic in particular. We also review the code snippets and verify that related and unrelated questions do not have any tags in common. 

Table \ref{tab:data_info} provides the attributes of the final datasets for each security tactic. The query parameter column lists the keywords used to query code snippets for each tactic. The sample size column provides the total number of related and unrelated code snippets in each dataset after the manual review process. In order to make sure that the number of samples for each security tactic is normalized, we create sample size categories for the number of related code snippets ranging between 25 and 200, with increments of 25. For example, the number of related code snippets found for SHA512 was 56, however only 50 randomly selected code snippets are used. Similarly, the number of code snippets for SHA256 was 139, but only 125 randomly selected samples are included. The training data for each security tactic is generated by merging an equal number of tactic related and unrelated code-snippets to have a balanced class distribution. For instance, for SHA512 and SHA256 security controls, there are 50 and 125 related samples respectively. Since the number of unrelated samples are equal to related samples, the total number of samples for SHA512 and SHA256 is 100 and 250, respectively. The columns S1, S2, S3 and S4 represent the percentage of samples that are represented with 1, 2, 3, and 4 BERT input sequences. The SMax column shows the number of sequences that the longest code snippet in a dataset is divided into.

\subsection{Pre-processing Code Snippets and Fine-tuning BERT}
\label{tactic_classification}
The process pipeline used to pre-process code snippets, fine-tune BERT, and classify its output sequences is shown in Figure \ref{fig:visual_overview}. Once the training data is pulled and manually reviewed for each security tactic, we perform a number of standard NLP pre-processing steps to make sure that all code snippets are tokenized, normalized and cleaned properly before BERT models are fine-tuned. After code snippets are tokenized, we split all camel case and snake case words, and convert all code-snippets to lower case for normalization. To clean the code snippets, all punctuations, numbers, short words less than two characters, and long words with more than 25 characters are removed. We also remove keywords of the Java language, because these keywords are common for all security controls and do not help to understand the security functionality of a code snippet. 

\begin{figure}[!h]
\centering
\includegraphics[width=\columnwidth]{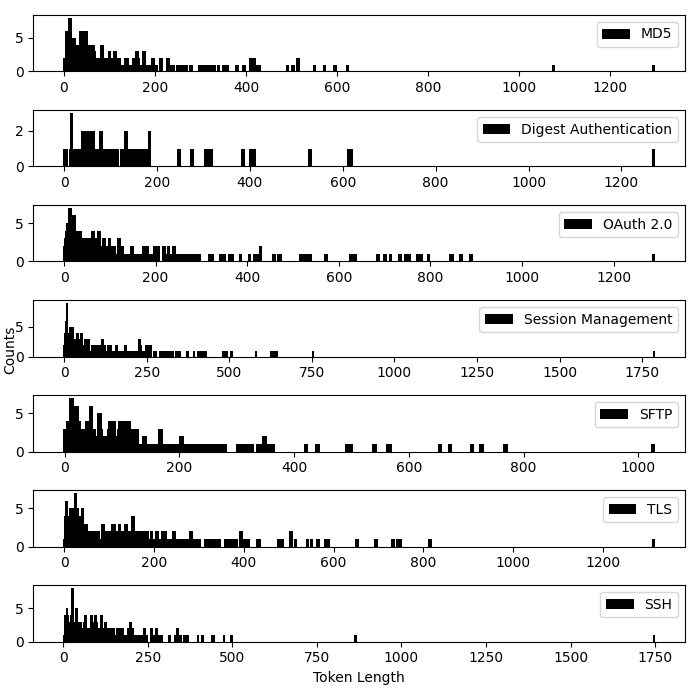}
\caption{The distribution of the token lengths of code snippets for security controls that have 3 or 4 in the SMax column in Table \ref{tab:data_info}}
\label{fig:fig_length_distr}
\end{figure}

The distribution of the length of the code snippets might be different for each tactic data set. The BERT algorithm can process input sequences with a maximum length of 510 tokens, and a special treatment is needed when the number of tokens is greater than 510 to make sure all tokens in the input sequences are considered. Figure \ref{fig:fig_length_distr} shows the distribution of the number of tokens for the security controls that include relatively larger code snippets. Although the majority of the observed token lengths are less than 510, these security controls still include some code snippets that have more than 1000 or even 1500 tokens. In order to make sure that the methodology used in this paper is applicable to larger code snippets as well, we apply mean pooling and make sure that calculated means represent the average pooled representation of the entire input sequence in a code snippet.

Out of the maximum 512 input tokens that can be accepted by BERT, two tokens are reserved for special tokens \ie CLS and SEP. Therefore, an input sequence can only be of 510 tokens, and extra tokens after the first 510 tokens are ignored by BERT. We divide large input sequences into multiple sequences of 510 tokens. If $N$ is the number of tokens in a given code-snippet, then we divide the code-snippet into $m=N / 510$ sequences. We get the BERT representation of the $m$ sequences and use mean pooling to combine them to get the final representation for the entire input code-snippet.

\subsection{Tactic Classifiers}
\label{tactic_classification_methods}
We fine-tune the \textit{bert-base-uncased} model \cite{bert-base-uncased} with the tactic datasets explained in Section \ref{tactic_dataset}. The base model has 12 layers of transformer encoder blocks, 768 hidden units and 12 self-attention heads. It is fine-tuned for code-snippet classification and used to predict whether a code snippet is related to a particular security tactic or not. Each BERT model is trained for 10 \emph{epochs} with a \emph{batch size} of 16, \emph{learning rate} of 2e-5, and a maximum sequence length of 512. A classifier with one hidden layer (including 32 neurons) and an output layer is added on top of BERT and the output vector corresponding to the [CLS] token is used to classify each input sequence. This vector is the average pooled representation of the entire input sequence and hence can be used for classification. The weighted sum of the output vector corresponding to the [CLS] token is calculated and passed through the classifier and the Softmax function is used to get the final probability indicating whether a tactic is implemented or not. During fine-tuning, all BERT parameters along with the classification layer parameters ($W$ and $B$) are updated to maximize the log-probability for the predicted tactic. 
 

To classify security tactics with an alternative method, we use the Tactic Detector (TD) introduced in our previous work \cite{Mirakhorli2016}. TD is a customized version of the information retrieval method developed to automate the detection and classification of non-functional requirements (NFRs) \cite{Cleland_2006_Tactic_Classifier}. In our previous work, using code snippets derived from 50 open source systems, we trained seven off-the-shelf classifiers including Support Vector Machine (SVM), C.45 Decision Tree (DT), Bayesian Logistic Regression (BLR), AdaBoost, rule learning with SLIPPER, Bagging, and Voting. We used these classifiers and the TD method to detect tactics in the Hadoop Distributed File System. The customized TD method outperformed other classifiers and detected tactic types with recall rates greater than 0.7. Leveraging findings of our prior work, this paper included TD as an alternative tactic detection method to classify the the same set of pre-processed code snippets used for BERT.

\section{Experiments and Results}
\label{experiments_results}
This paper uses BERT and TD as tactic classification methods and designs three sets of experiments and a case study to evaluate the performance of the proposed methodology:
\begin{itemize}
    \item \textbf{Experiment 1 :} Related and unrelated code snippets of each tactic are used to predict whether a source file is tactic-related or not.
    \item \textbf{Experiment 2 :} Related and unrelated code snippets within the same category are merged to predict whether a code piece is tactic-related. We merge all controls under the yellow branches in the second level of the categorization tree in Figure \ref{fig:tactic_categorization_chart} \ie Data Integrity, Authenticate, Encrypt/Decrypt, Authorize and Secure Communication.
    \item \textbf{Experiment 3 :} This is a multi-label classification experiment, where the tactic names in Experiment 2 are used as class labels. Only related code snippets from each tactic (within the same category) are combined to form the datasets for this experiment.
    \item \textbf{Case Study :} To evaluate the performance of the proposed methodology on a real life project, we conduct a case study on a large scale open source system, {\texttt{Liferay portal}} (\href{https://github.com/liferay/liferay-portal}{https://github.com/liferay/liferay-portal}), that is implemented in the Java language.
\end{itemize}


\begin{table}[!h]
\centering
\caption{Precision (P), Recall (R) and F-Measure (F) values obtained with BERT and TD during Experiment 1. The Sample Size (SS) column gives the total number of training code snippets used for each tactic. }
\label{tab:experiment_1_results}

\begin{tabular}{p{2cm}p{0.5cm}p{0.5cm}p{0.5cm}p{0.5cm}p{0.5cm}p{0.5cm}p{0.5cm}}
\thickerhline
                      &             & \multicolumn{3}{c}{\textbf{BERT}}                                               & \multicolumn{3}{c}{\textbf{TD}}                                                 \\
                       \thickhline
\textbf{Tactic}       & \textbf{SS} & \multicolumn{1}{c}{\textbf{P}} & \multicolumn{1}{c}{\textbf{R}} & \textbf{F}    & \multicolumn{1}{c}{\textbf{P}} & \multicolumn{1}{c}{\textbf{R}} & \textbf{F}    \\
 \thickhline
SHA256                & 500         & 1                              & 0.99                           & \textbf{0.99} & 1.00                           & 0.93                           & 0.96          \\
SHA512                & 200         & 0.98                           & 0.98                           & 0.98          & 1.00                           & 0.96                           & 0.98          \\
MD5                   & 800         & 0.99                           & 0.99                           & \textbf{0.99} & 0.99                           & 0.98                           & 0.98          \\
Token Authentication  & 400         & \multicolumn{1}{c}{0.96}       & \multicolumn{1}{c}{0.96}       & \textbf{0.96} & 0.99                           & 0.84                           & 0.91          \\
Digest Authentication & 200         & 0.93                           & 0.96                           & \textbf{0.95} & 1.00                           & 0.82                           & 0.90          \\
LDAP                  & 800         & 1                              & 0.99                           & \textbf{0.99} & 1.00                           & 0.83                           & 0.90          \\
Kerberos              & 300         & 0.98                           & 1                              & \textbf{0.99} & 1.00                           & 0.72                           & 0.84          \\
AES                   & 800         & 0.96                           & 0.91                           & \textbf{0.94} & 0.98                           & 0.81                           & 0.88          \\
Blowfish              & 200         & 0.96                           & 0.96                           & 0.96          & 1.00                           & 0.98                           & \textbf{0.99} \\
RSA                   & 800         & 0.99                           & 0.99                           & \textbf{0.99} & 0.97                           & 0.95                           & 0.96          \\
3DES                  & 200         & 0.93                           & 1                              & 0.97          & 1.00                           & 0.96                           & \textbf{0.98} \\
OAuth 2.0             & 800         & 0.97                           & 0.99                           & \textbf{0.98} & 0.99                           & 0.91                           & 0.95          \\
Access Control List   & 300         & 0.96                           & 0.96                           & \textbf{0.96} & 1.00                           & 0.76                           & 0.86          \\
Session Management    & 700         & \multicolumn{1}{c}{0.97}       & \multicolumn{1}{c}{0.97}       & \textbf{0.97} & 0.99                           & 0.86                           & 0.92          \\
Input Validation      & 100         & 0.86                           & 1                              & \textbf{0.92} & 0.94                           & 0.68                           & 0.79          \\
SFTP                  & 800         & \multicolumn{1}{c}{0.99}       & \multicolumn{1}{c}{0.98}       & \textbf{0.98} & 1.00                           & 0.87                           & 0.93          \\
TLS                   & 800         & \multicolumn{1}{c}{0.98}       & \multicolumn{1}{c}{0.96}       & \textbf{0.97} & 1.00                           & 0.79                           & 0.88          \\
VPN                   & 100         & 0.86                           & 0.9                            & \textbf{0.88} & 0.94                           & 0.60                           & 0.73          \\
SSH                   & 800         & \multicolumn{1}{c}{0.97}       & \multicolumn{1}{c}{0.98}       & \textbf{0.97} & 0.99                           & 0.85                           & 0.91          \\
Audit trail           & 400         & 0.98                           & 0.99                           & \textbf{0.99} & 0.99                           & 0.79                           & 0.87         \\
\thickerhline
\end{tabular}
\end{table}

Precision, Recall and F-Measure metrics are used to evaluate the performance of each model. The Precision metric represents the fraction of correctly identified tactic-related code snippets within all tactic-related (positive) predictions. Similarly, Recall represents the fraction of correctly identified tactic-related code snippets. We calculate the harmonic mean of Precision and Recall to calculate F-Measure by
\begin{equation}
F-Measure = 2.\frac{Precision.Recall}{Precision+Recall}
\end{equation}
All experiments are run with a stratified 10-fold cross validation approach. Each tactic dataset is split into 10 folds such that the distribution of the classes is the same in each fold. Each method is trained with nine folds and tested on the remaining fold. Each fold is used as test set, and at the end, the average of test metrics obtained from 10 folds is used to evaluate overall model performance. 

\subsection{Experiment 1: Binary Tactic Classification}
\label{experiment_1}
In an enterprise level large software project, it could be very helpful to have an automated system to identify the files, classes or even packages that implement a security tactic, without going through a manual code review process. Given a specific security tactic, being able to generate the list of files or classes where the tactic is implemented gives software professionals a chance to review those files or classes quickly and identify security related architectural weaknesses.  

\begin{figure}[h!]
\centering
\includegraphics[width=\columnwidth]{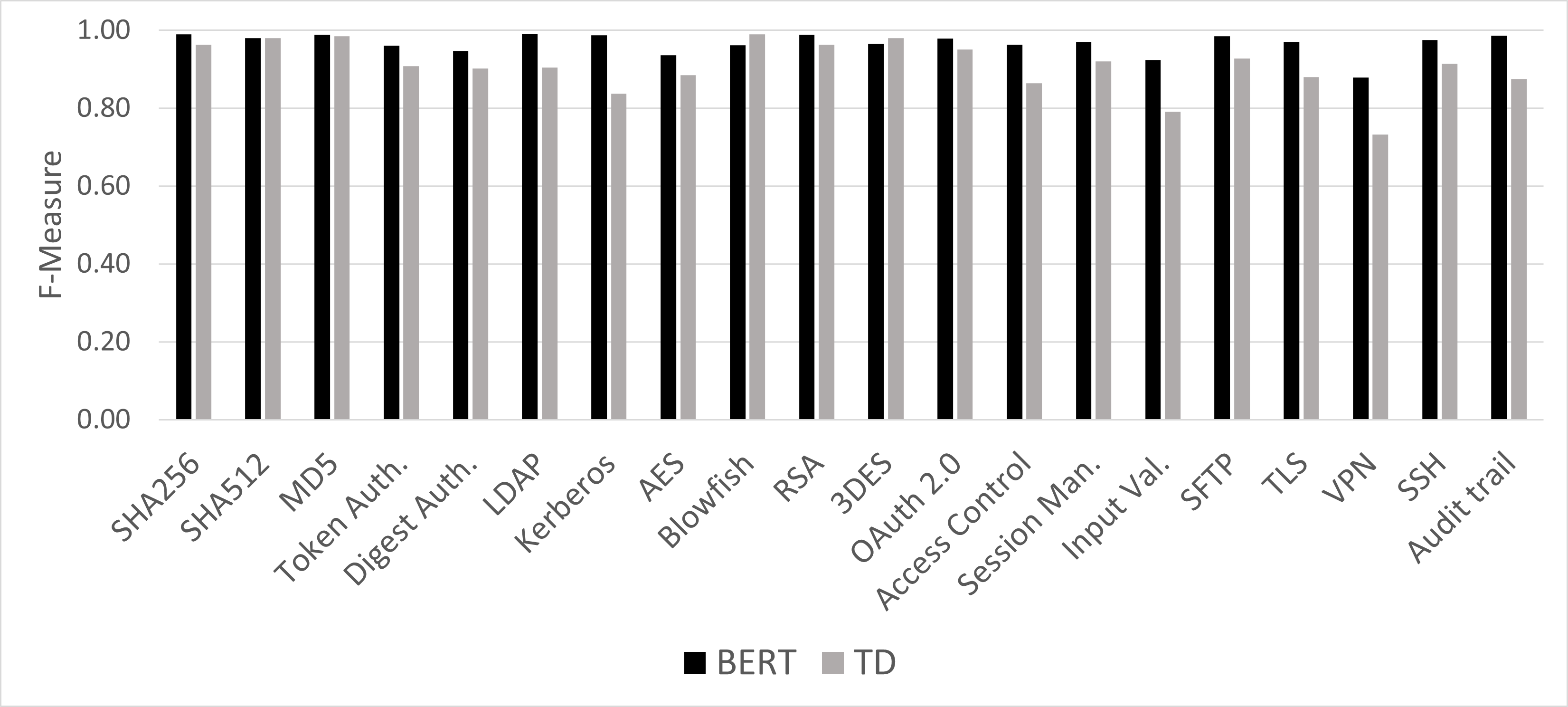}
\caption{A comparison of the $F$-Measure values for BERT and TD in Experiment 1.}
\label{fig:fig_exp1_bert_vs_td25}
\end{figure}

\begin{figure}[h!]
\centering
\includegraphics[width=\columnwidth]{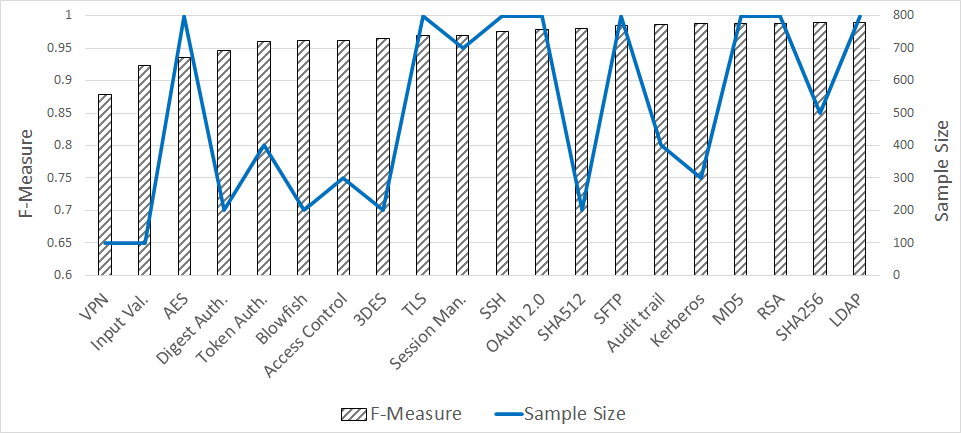}
\caption{$F$-Measure values obtained with BERT and the training data sample sizes in Experiment 1.}
\label{fig:fig_exp1}
\end{figure}

The goal of this experiment is to determine how well the models are able to differentiate between the tactic related and unrelated code snippets for each tactic. Table \ref{tab:experiment_1_results} shows the Precision, Recall and F-Measure scores obtained with BERT and TD for each tactic. For BERT, the F-Measure values range from 0.88 to 0.99 and the average F-Measure for all controls is 0.97. Most models are able to effectively capture the distinctive features in the code snippets, even with small sample sizes. The F-Measure values of TD range between 0.73 and 0.99 and the average F-Measure for all controls is 0.91. Figure \ref{fig:fig_exp1_bert_vs_td25} shows the comparison of F-Measure values obtained with BERT and TD. In general, we obtain better F-Measure values with BERT, except SHA512, 3DES and Blowfish that have relatively smaller training data sizes. \textit{Mann–Whitney U test} is used to compare the F-Measure values obtained with BERT and TD and the result shows that the difference is significant with a $p-$value of 0.001. Figure \ref{fig:fig_exp1} shows that except AES, lower F-Measure values are obtained with BERT when the number of training samples is small. The lowest F-Measure values are observed for VPN and Input Validation are 0.88 and 0.92, respectively, and the sample size for these controls is 100. 

Similar to the prior work by Ezen-Can\cite{ayse2020_bert}, the results of Experiment 1 indicate that in general lower F-Measure values are obtained with BERT when the training data set size is small. 

\subsection{Experiment 2: Binary Tactic Classification within Category}
\label{experiment_2}
A large software project that have thousands of files, could potentially have hundreds of files or classes that implement a variety of security controls. In such a case, being able to identify the files or classes that implement one or more (potentially different) security controls is important to help code reviewers focus on the tactic related code pieces while identifying architectural flaws.

\begin{table}[h!]
\centering
\caption{Precision, Recall and F-Measure values for each category in Experiment 2.}
\label{tab:experiment_2_results}

\begin{tabular}{p{2.3cm}p{0.5cm}p{0.4cm}p{0.4cm}p{0.4cm}p{0.4cm}p{0.4cm}p{0.4cm}}
\thickerhline
                         &             & \multicolumn{3}{c}{\textbf{BERT}}       &
                         \multicolumn{3}{c}{\textbf{TD}}    \\
                         \thickhline
\textbf{Tactic Category} & \textbf{SS} & \textbf{P} & \textbf{R} & \textbf{F}    & \textbf{P} & \textbf{R} & \textbf{F} \\
\thickhline
Data Integrity           & 1500        & 0.98       & 0.98       & \textbf{0.98} & 1.00       & 0.82       & 0.90       \\
Authenticate             & 1700        & 0.96       & 0.97       & \textbf{0.96} & 1.00       & 0.69       & 0.82       \\
Encrypt / Decrypt        & 2000        & 0.97       & 0.92       & \textbf{0.94} & 0.97       & 0.89       & 0.93       \\
Authorize                & 2600        & 0.96       & 0.98       & \textbf{0.97} & 0.99       & 0.64       & 0.78       \\
Secure Communication     & 2500        & 0.96       & 0.97       & \textbf{0.96} & 0.99       & 0.77       & 0.87 \\
\thickerhline
\end{tabular}
\end{table}

During this experiment, a category dataset is created by combining the tactic related and unrelated code-snippets of all security controls within the same category. The models are trained with the combined code snippets of each category, and 10 folds stratified cross validation approach is used to determine how well the models differentiate security related and unrelated code-snippets. 

\begin{figure}[!h]
\centering
\includegraphics[width=\columnwidth]{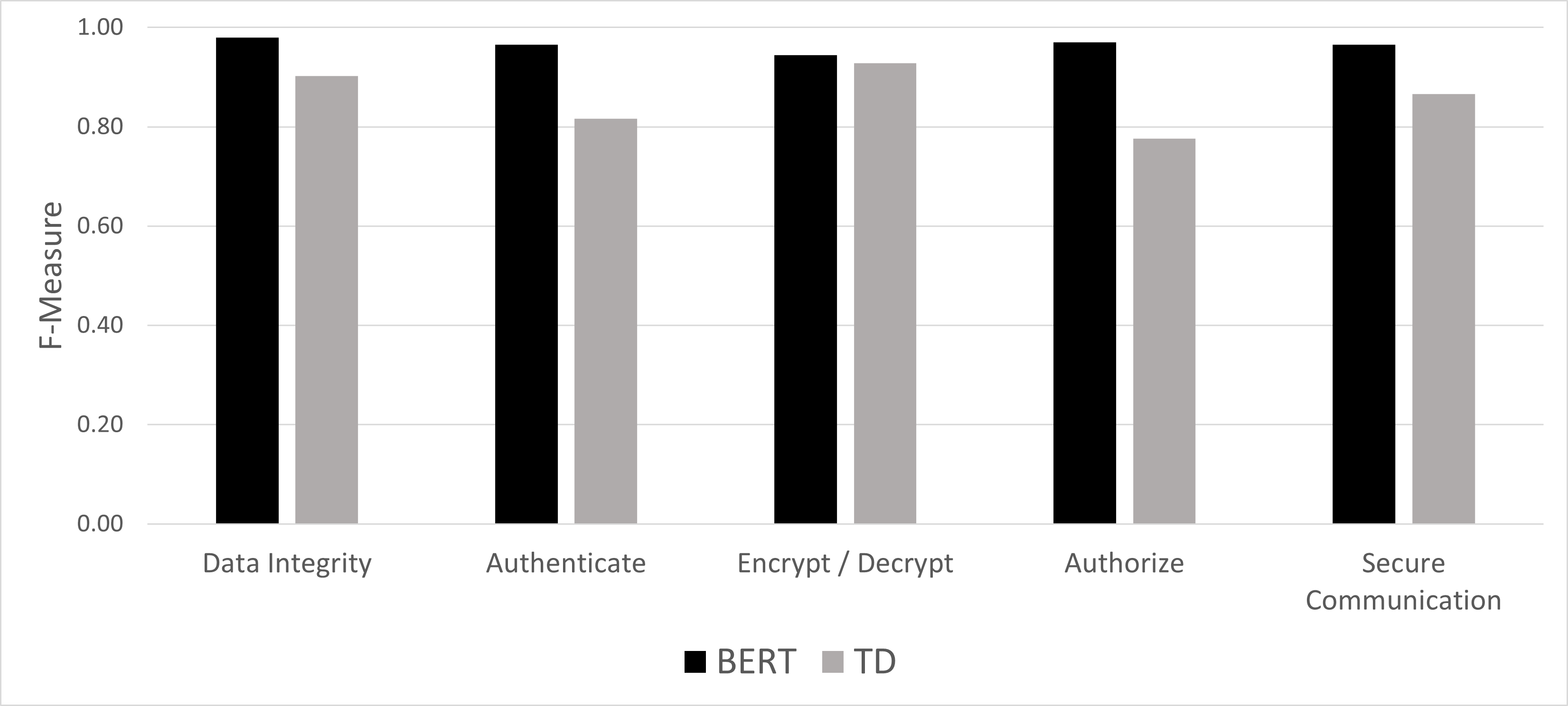}
\caption{A comparison of the $F$-Measure values for BERT and TD in Experiment 2.}
\label{fig:fig_exp2_bert_vs_td25}
\end{figure}

Table \ref{tab:experiment_2_results} shows the Precision, Recall and F-Measure scores for each tactic category for BERT and TD methods, and Figure \ref{fig:fig_exp2_bert_vs_td25} shows the comparison of their F-Measure values. BERT models perform quite well with F-Measure values ranging from 0.94 to 0.98, where the average F-Measure for all categories is 0.96. Similarly, TD achieves F-Measure values ranging between 0.78 and 0.93 with an average F-Measure of 0.86. One reason for observing such higher F-Measure values (compared to Experiment 1) is that related code-snippets of the security controls within the same category have similar implementations. For example, the security controls under Secure Communication \ie SFTP, SSH and TLS all include similar steps to establish a connection and then interact with it. SFTP and SSH are both implemented using the Java Secure Channel framework. Similarly, all security controls under the Encrypt Decrypt category are implemented using the Java Cryptography Extension (JCE) framework. Moreover, Digest Authentication, Kerberos and Token Based Authentication in the Authenticate category are all using the Spring Security framework. 

\subsection{Experiment 3: Multi-label Tactic Classification}
Although it is more challenging compared to the cases discussed in Sections \ref{experiment_1} and \ref{experiment_2}, given a list of security controls and a large software project repository, it could be quite helpful to be able to identify which security tactic is implemented in the files, classes or packages in the repository. That could help to pinpoint the code pieces that implement the security controls of interest and fix architectural problems if any.

\begin{table*}[!h]
\centering
 \caption{Precision (P), Recall (R) and F-Measure (F) values obtained with BERT and TD during Experiment 3. The Sample Size (SS) column provides the total number of training code snippets for each tactic in each tactic category.}
    \label{tab:experiment_3_up_sampled_results}
\begin{tabular}{p{3cm}p{3.5cm}p{0.5cm}p{0.5cm}p{0.5cm}p{0.5cm}p{0.5cm}p{0.5cm}p{0.5cm}}
\thickerhline
                                               &                           & \multicolumn{1}{l}{} & \multicolumn{3}{c}{\textbf{BERT}}                                                                & \multicolumn{3}{c}{\textbf{TD}}         \\
\textbf{Tactic Category}                       & \textbf{Tactic}           & \textbf{SS}          & \multicolumn{1}{c}{\textbf{P}} & \multicolumn{1}{c}{\textbf{R}} & \multicolumn{1}{c}{\textbf{F}} & \textbf{P} & \textbf{R} & \textbf{F}    \\
\thickhline
\multirow{3}{*}{\textbf{Data Integrity}}       & SHA512                    & 200                  & 0.9                            & 0.97                           & \textbf{0.93}                  & 0.67                           & 0.03                           & 0.06          \\
                                               & SHA256                    & 200                  & 0.96                           & 0.9                            & \textbf{0.93}                  & 0.58                           & 0.73                           & 0.64          \\
                                               & MD5                       & 200                  & 1                              & 0.99                           & \textbf{0.99}                  & 0.60                           & 0.98                           & 0.74          \\
\multirow{4}{*}{\textbf{Authenticate}}         & Token Base Authentication & 200                  & 0.79                           & 0.82                           & 0.8                            & 0.80                           & 0.86                           & \textbf{0.83} \\
                                               & Digest Authentication     & 200                  & 0.79                           & 0.77                           & 0.78                           & 0.89                           & 0.87                           & \textbf{0.88} \\
                                               & Kerberos                  & 200                  & 0.81                           & 0.84                           & \textbf{0.82}                  & 0.92                           & 0.71                           & 0.80          \\
                                               & LDAP                      & 200                  & 0.9                            & 0.92                           & \textbf{0.91}                  & 0.90                           & 0.87                           & 0.88          \\
\multirow{4}{*}{\textbf{Encrypt / Decrypte}}   & AES                       & 200                  & 0.78                           & 0.76                           & 0.77                  & 0.79                           & 0.75                           & 0.77          \\
                                               & Blowfish                  & 200                  & 0.89                           & 0.99                           & 0.94                           & 0.97                           & 0.94                           & \textbf{0.96} \\
                                               & RSA                       & 200                  & 0.9                            & 0.95                           & \textbf{0.93}                  & 0.85                           & 0.88                           & 0.87          \\
                                               & 3DES                      & 200                  & 0.95                           & 0.9                            & \textbf{0.93}                  & 0.96                           & 0.84                           & 0.89          \\
\multirow{4}{*}{\textbf{Authorize}}            & LDAP                      & 200                  & 0.97                           & 0.96                           & \textbf{0.97}                  & 0.98                           & 0.89                           & 0.93          \\
                                               & Oauth 2.0                 & 200                  & 0.92                           & 0.97                           & \textbf{0.94}                  & 0.88                           & 0.92                           & 0.90          \\
                                               & ACL                       & 200                  & 1                              & 0.98                           & \textbf{0.99}                  & 1.00                           & 0.82                           & 0.90          \\
                                               & Session Management        & 200                  & 0.98                           & 0.95                           & \textbf{0.96}                  & 0.91                           & 0.92                           & 0.91          \\
\multirow{4}{*}{\textbf{Secure Communication}} & SFTP                      & 200                  & 0.86                           & 0.95                           & \textbf{0.9}                   & 0.81                           & 0.83                           & 0.82          \\
                                               & TLS                       & 200                  & 0.96                           & 0.96                           & \textbf{0.96}                  & 0.91                           & 0.75                           & 0.82          \\
                                               & VPN                       & 200                  & 0.99                           & 1                              & \textbf{0.99}                  & 1.00                           & 0.65                           & 0.79          \\
                                               & SSH                       & 200                  & 0.95                           & 0.83                           & \textbf{0.89}                  & 0.77                           & 0.69                           & 0.72    \\
                                               \thickerhline
\end{tabular}
\end{table*}

Related code snippets within each category (of Experiment 2) are merged to form a dataset for multi-class classification where tactic names are used as class labels. The goal of this experiment is to measure the performance of the fine-tuned BERT models and TD while differentiating controls within the same category. There are two main challenges for this experiment. First, security controls within the same category might use common libraries or frameworks and have quite similar implementations. Therefore, it is more challenging to distinguish code snippets that have very similar implementations of the security controls in the same category. Second, the number of samples for each security tactic within each category is different and the distribution of the classes is imbalanced. For example, under the Encrypt Decrypt tactic category, AES, Blowfish, RSA, and 3DES have samples sizes of 200, 50, 200, and 50, respectively. Similarly, the classes are imbalanced for other categories as well. Models trained with imbalanced data sets might have a tendency to ignore the minority classes and over-fit on the majority ones. Some common methods\cite{a_study_of_sampling_techniques} to balance class distributions are random under sampling, SMOTE\cite{chawla2011smote}, and random oversampling. Random under sampling could lead to a performance loss due to smaller overall sample sizes. It is hard to use SMOTE, because generating syntactically and semantically valid code-snippets is not a trivial task. Therefore, we perform random oversampling for the minority classes in each category to have a balanced class distribution for the security controls within the same category. 

\begin{figure}[!h]
\centering
\includegraphics[width=\columnwidth]{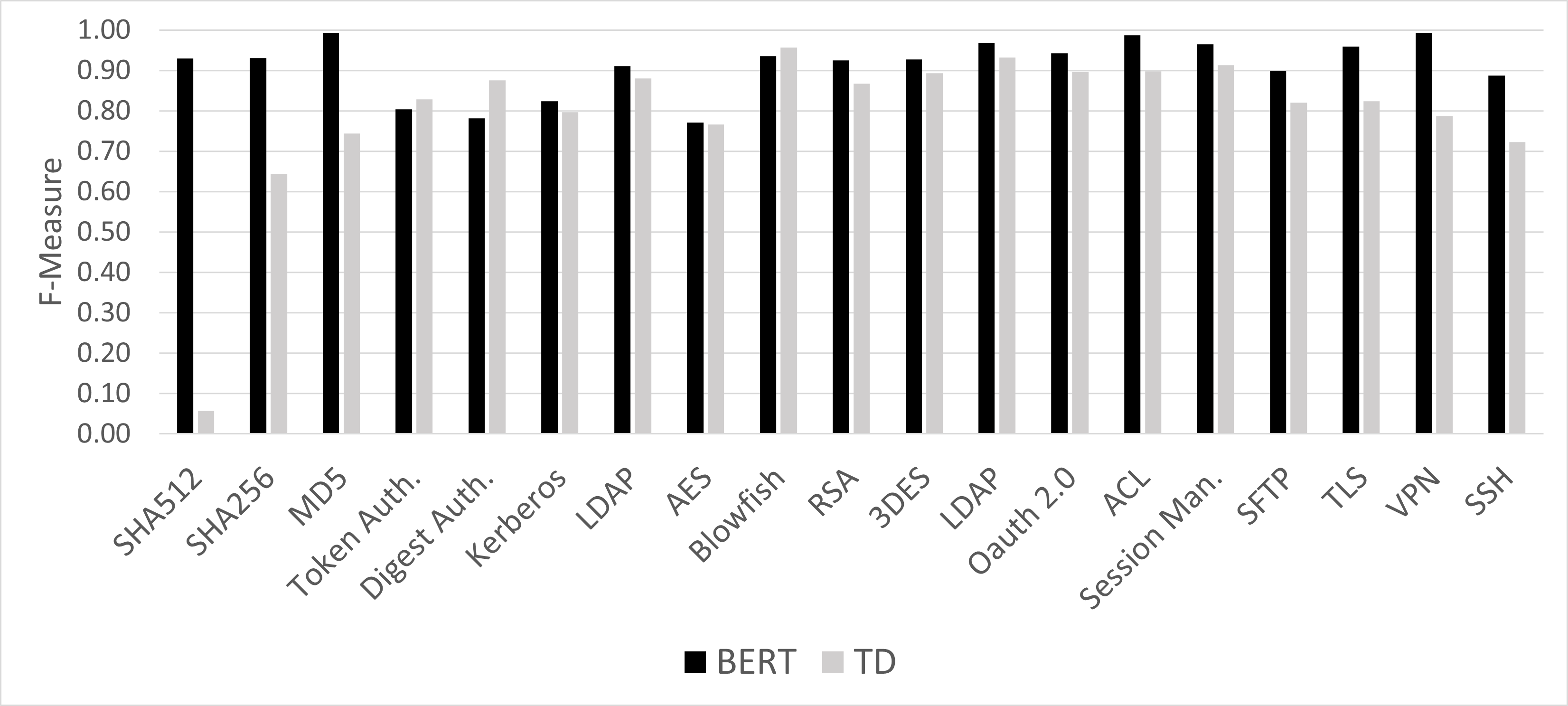}
\caption{A comparison of the $F$-Measure values of BERT and TD in Experiment 3.}
\label{fig:exp3_fm}
\end{figure}

Table \ref{tab:experiment_3_up_sampled_results} shows Precision, Recall and F-Measure values for each security tactic with BERT and TD. The average F-Measure scores of BERT for the Data Integrity, Authenticate, Encrypt / Decrypt, Authorize, and Secure Communication tactic categories are 0.95, 0.83, 0.89, 0.97, and 0.93, respectively and the overall average for the experiment is 0.91. Figure \ref{fig:exp3_fm} shows the comparison of the F-Measure values obtained for each tactic with BERT and TD. Most F-Measure values obtained with BERT are above 0.9, and BERT outperforms TD during this experiment. \textit{Mann–Whitney U test} test is used to run statistical significance tests on the F-Measure values obtained with BERT and TD and the result shows that the difference between the two sets of F-Measure values is significant with a $p-$value of 0.002.

\begin{figure}[!h]
\centering
\includegraphics[width=0.9\columnwidth]{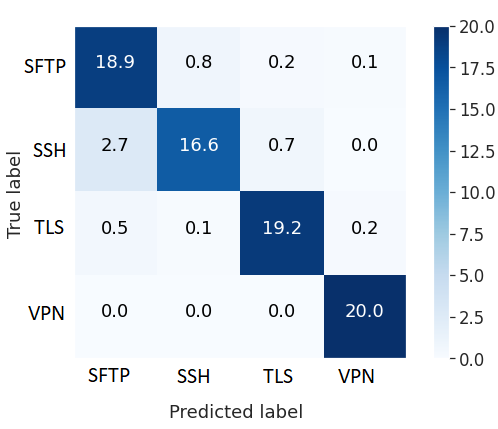}
\caption{The average of the confusion matrix entries obtained in each fold while classifying tactics in Secure Communication category.}
\label{fig:exp3_sc_cm}
\end{figure}

The results also suggest that the trained models are not that successful while differentiating security controls with very similar implementations. Security controls with lower performance scores tend to have very similar implementations with other security controls in the same category. For example, both SFTP and SSH use the Java Secure Channel framework and have similar implementations, therefore their F-Measure values are lower, compared to the F-Measure values obtained for TLS and VPN. As an example, the confusion matrix of the 10 folds BERT multi-label classifier for the Secure Communication category is shown in Figure \ref{fig:exp3_sc_cm}. The values in the confusion matrix show that SFTP is misclassified as SSH, and similarly SSH is misclassified as SFTP due to the similarity of their implementations. TLS and VPN are more distinguishable from other security controls compared to SFTP and SSH. 

\subsection{Experiment 4: Case Study}
To evaluate the performance of BERT and TD in a real-life large-scale project, a case study was conducted on a popular open source project, Liferay\footnote{https://github.com/liferay/liferay-portal} portal. Liferay is a well-known and widely used portal, mostly written in Java language. Liferay was chosen for the case study due to its vast application as a popular platform in various domains, including B2B online revenue, customer service, resource management, and system integration. 
A team of three experts studied over 43K of java files in the Liferay project to identify tactic-related files. During this manual inspection, 210 instances of LDAP, 96 instances of MD5, 84 instances of SHA, and 389 instances of OAUTH2 related files were found. The team labeled all files as positive or negative based on their relationship with the mentioned tactics. For instance, 210 files were labeled as LDAP, while the remaining 42790 files were labeled as unrelated to LDAP. To create a test dataset that includes a fair number of related and unrelated test cases, we followed the 20\% - 80\% ratio between related and unrelated test cases. In that regard, from the files labeled as unrelated to LDAP, the team randomly selected 840 files to be included in the case study. Table \ref{tab:experiment_4_results} presents the sample size of test cases for each tactic under the \textbf{SS} column.  

During the case study, we investigated the performance of both the Tactic Detector, and the Bert model. We first trained a binary model for each tactic using its data set summarized in Table \ref{tab:data_info}, where 200 files are labeled related (positive) for MD5, LDAP, and OAuth 2.0 and 150 files are marked positive for SHA. Then, we used the manually labeled and reviewed Liferay Java files as the test data set, to measure the performance of each trained model. Before predicting whether a tactic is implemented or used in a file, we performed a set of pre-processing steps. All Liferay Java files include the same \textit{comment} block expressing copyright-related information at the beginning of the file. We removed these comment blocks since they are common to all files and do not provide file-specific information. Other pre-processing steps included tokenizing each file, removing punctuation and Java language keywords, splitting each token based on camel case convention, removing stop words and numbers, and finally, removing lengthy (over 50 characters) and short (less than 2 characters) tokens.  

\begin{table*}[h!]
    \caption{Precision, Recall and F-Measure values per tactic for BERT and TD in Experiment 4.}
    \label{tab:experiment_4_results}
\centering
\begin{tabular}{p{1cm} p{1.2cm}  p{1.4cm} p{1cm} p{.8cm} p{1.4cm} p{1cm} p{.8cm} p{1.4cm} }
 \thickerhline
  \multirow{2}{4em}{\textbf{Tactic}}& \multirow{2}{4em}{\textbf{Label}}& \multirow{2}{4em}{\textbf{SS}} & \multicolumn{3}{c}{\textbf{BERT}}&
  \multicolumn{3}{c}{\textbf{TD}}\\
  
     & & & \textbf{P}&\textbf{R}&\textbf{F}&\textbf{P}&\textbf{R}&\textbf{F}\\

  \thickhline
  \multirow{2}{4em}{LDAP} & 
  LDAP& 210 & 0.49&	1&	0.65& 0.73& 0.79	&	\textbf{0.76}\\
  & Unrelated& 840 & 1&	0.74&	0.85& 0.95&	0.93&	\textbf{0.94} \\
    \hline
  \multirow{2}{4em}{MD5} & 
  MD5& 96 &0.51&	0.84&	0.64& 0.99 & 0.75 & \textbf{0.85}\\
  & Unrelated & 384 & 0.95&	0.80&	0.87& 0.94 & 1.00 & \textbf{0.97}\\
    \hline
  \multirow{2}{4em}{SHA} & 
  SHA& 84 & 0.83 &	0.65&	\textbf{0.73}& 0.92 & 0.54 & 0.68\\
  & Unrelated & 336 & 0.92&	0.97&	\textbf{0.94}& 0.89 & 0.99 & \textbf{0.94}\\
    \hline
  \multirow{2}{4em}{OAUTH2} & 
  OAUTH2& 389 & 0.25 &	0.92&	0.39& 0.73 & 0.79 & \textbf{0.76}\\
  & Unrelated & 1556 & 0.94&	0.30&	0.45& 0.95 & 0.93 & \textbf{0.94}\\
  
\thickerhline
\end{tabular}
\end{table*}

Table~\ref{tab:experiment_4_results} demonstrates the results of the experiments during the case study. Based on the recalls shown in this table, BERT model was able to correctly identify 100\%, 84\%, 65\%, and 92\% of LDAP, MD5, SHA, and OAUTH2 cases respectively. TD, on the other side, identified 79\%, 75\%, 54\%, and 79\% of LDAP, MD5, SHA, and OAUTH2 tactic-related test samples. BERT outperforms TD when identifying tactic-related cases,  however based on the recalls of unrelated cases shown in Table \ref{tab:experiment_4_results}, TD identifies a higher percentage of unrelated cases. Another measure provided in Table \ref{tab:experiment_4_results} is the precision of the predictions. Although BERT results show a slightly better (overall) precision for unrelated cases, TD has a significantly better precision for the tactic-related test cases.  


We investigated the reasons for relatively high rate of false-positives in tactic-related predictions by BERT model. We particularly identified three main drivers. First, the imbalanced nature of the test data could be a driver for high false positives. For instance, there are only 210 LDAP samples whereas the number of unrelated code snippets to LDAP is 840 samples. Based on this distribution of the data, even if the model correctly identifies all the LDAP samples (210 cases) and incorrectly identifies only 20.12\% of the unrelated samples (169 cases), the precision of LDAP would be 55\%. Second, the existence of security-related terms and keywords in the files that are labeled as \textit{Unrelated} may result in false-positive predictions. For instance, we found \quotes{User,} \quotes{Permission,} \quotes{Access,} and some other security-related terms in the files classified as false-positive. If a file implements a security related feature (even if it is not exactly the tactic under study), there is a chance that the model labels that file as a tactic implementation. Lastly, BERT model needs a large training dataset to be able to be fine tuned \cite{gong2019efficient}. Therefore, training such a model by a dataset of 400 training instances (LDAP, MD5, OAUTH2) or even fewer (SHA), can result in a model that does not perform well when it comes to test cases.

\begin{figure}[!h]
\centering
\includegraphics[width=\columnwidth]{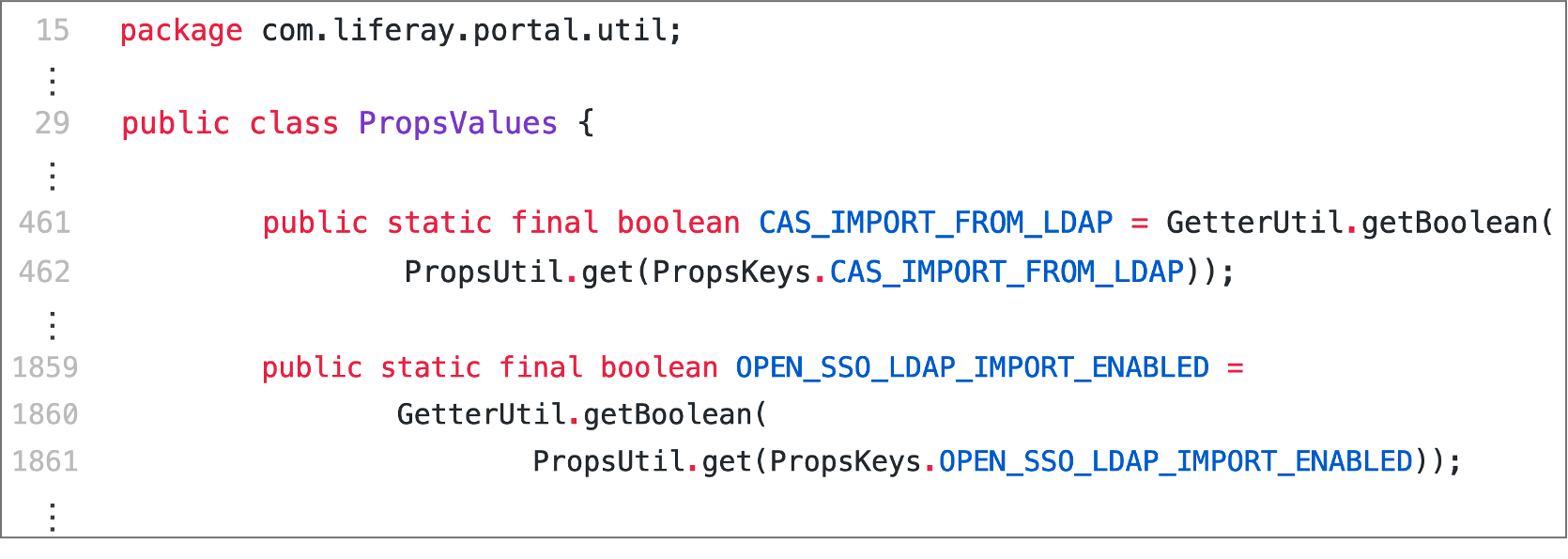}
\caption{Part of a Java file in Liferay implementation module which normally needs manual inspection to be identified as Unrelated to the LDAP tactic.}
\label{fig:exp4_code_snippet}
\end{figure}

In general, the F-measure of TD approach shown in Table \ref{tab:experiment_4_results} is relatively better compared to the BERT. However, considering the lower amount of false-negatives for tactic-related predictions of BERT compared to TD, we believe that BERT can filter out source files that do not implement a specific tactic with a higher confidence compared to the TD, thereby decrease the time needed for manual code inspection. The contribution of the BERT could be more significant when a software project is composed of large number of files and an automated method is needed to conservatively rule out the modules or files that do not implement a specific security tactic. 
For example, Figure \ref{fig:exp4_code_snippet} shows part of a Java file in the Liferay portal which defines some static values to be used by other classes, but does not include a tactic implementation. For instance, at line 461 and 462, the programmer defines a Boolean variable \textsc{CAS\_IMPORT\_FROM\_LDAP} and assigns a value to it. Furthermore, through the lines 1859 to 1861, another Boolean variable \textsc{OPEN\_SSO\_LDAP\_IMPORT\_ENABLED} is defined and initialized. In both the cases, despite the fact that these lines include the term \textit{LDAP}, the file does not actually use or implement the LDAP tactic. During the case study, this file was correctly identified as \textit{Unrelated} to the LDAP tactic. However, it normally takes time and effort for a programmer to manually inspect such files and decide whether a certain tactic is implemented or not. In addition, this manual process could be error prone. 

\begin{figure}[!h]
\centering
\includegraphics[width=\columnwidth]{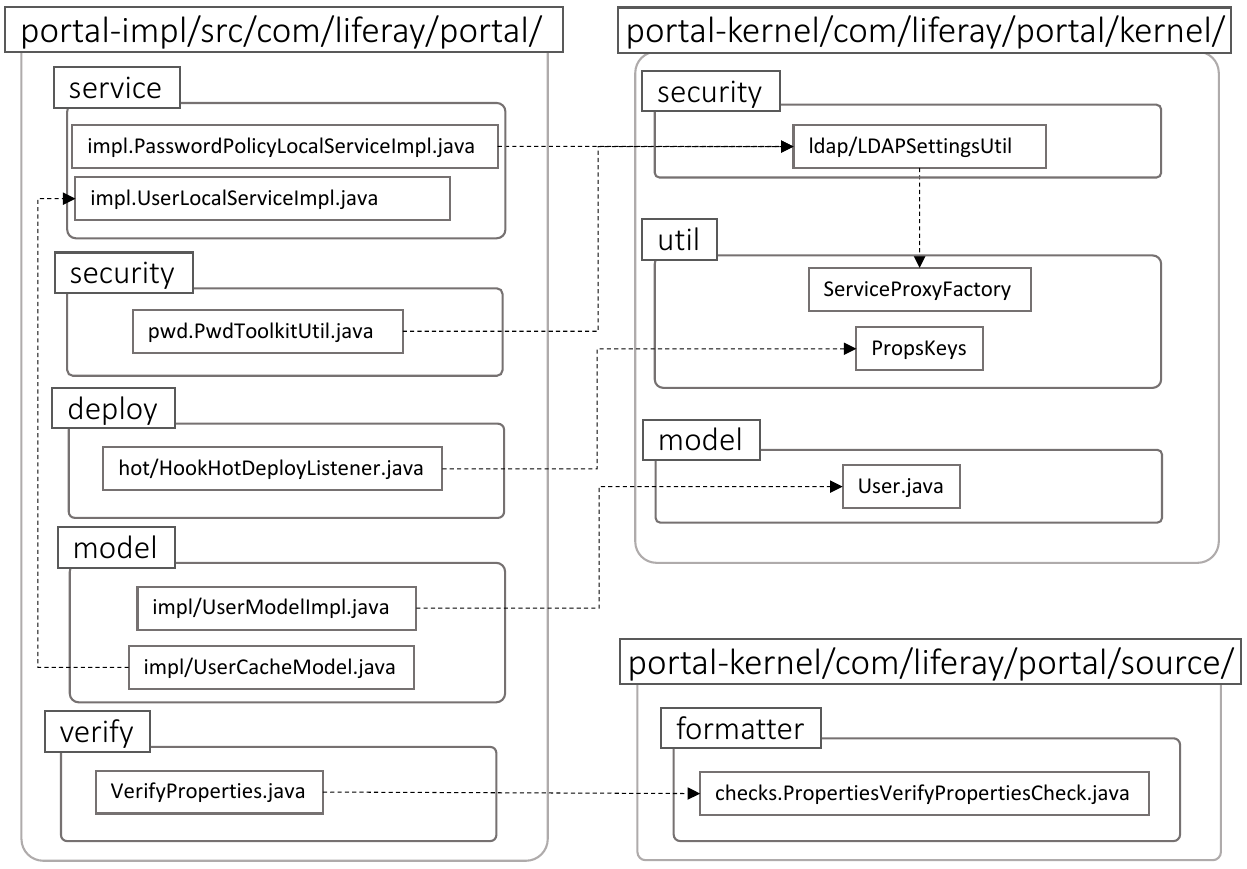}
\caption{LDAP related files detected by the BERT model (left) and their dependencies to other files in different folders and packages.}
\label{fig:exp4_ldap_liferay_files}
\end{figure}

Due to its high recall for tactic-related instances, the BERT approach presented by this work can also enable  architectural diagrams generators to show tactic related modules/files and their dependencies. Figure \ref{fig:exp4_ldap_liferay_files} shows part of the LDAP related folders and files in the Liferay project and their dependencies to other modules and files. The files shown on the left hand side are correctly identified as LDAP related and have dependencies to other services and utilities in the Liferay project shown on the right hand side. In projects with very large number of files, packages, and modules, manually tracing module dependencies and judging whether a file implements or uses a tactic is a tedious and risky task. Leveraging the automated approach proposed by this paper, files that do not implement or use a tactic could be automatically ruled out from the manual inspection process, and that could lead to a significant reduction in the time and effort needed to identify tactic-related files.

\subsection{Use Cases}
\label{use_cases}
The results of this work have a couple of practical implications for the security and privacy of software projects in real life:

\noindent \circled{1} \textbf{Identification of Security Tactics in Source Code: }
This work can be used to pinpoint parts of a software project that implement a security tactic and help to identify existing architectural flaws earlier. Fine tuned BERT models or TD can be used to identify whether a given security tactic (or either one of a group of security controls) is implemented in the files or classes of a software project in Java language. The trained models can further help to identify the type of security controls that are implemented in a project.

\noindent \circled{2} \noindent \textbf{Security Requirement Tracing: } 
Given a set of security requirements, the models can be used to trace these requirements and check if they are properly implemented. 

\noindent \circled{3} \textbf{Visualization of Tactic Related Modules and their Dependencies: } 
The output of the proposed method can be used to produce visualizations for the software architecture to highlight modules or packages where security controls are used or implemented and modules that they have a dependency with.

\section{Threats to Validity}
\label{Threats_to_Validity}
Perry \etal state that external, construct and internal validity are three types of validity threats that should be addressed in research studies\cite{Perry2000}. In this section, we briefly discuss the threats that may affect the results of this study, and the measures taken to mitigate them.

External validity evaluates whether the results of a study can be generalized or not and is mostly related to the selection or construction of the datasets used. All tactic related and unrelated code snippets were pulled from the public API of the Stack Overflow platform and reviewed carefully to make sure that all included code pieces were labelled correctly, with a consistent approach. However, it is very difficult to argue that all possible implementations of all security controls are covered by the code snippets in our data sets.  

Construct validity threats can occur if a study has errors in the experiment methodology and the tests do not measure what is claimed properly. To mitigate construct validity threats, we use code snippets committed by the diverse user base of the Stack Overflow platform, rather than including code samples from a limited number of software projects. Furthermore, we apply 10 folds cross validation to make sure that we report consistent results for data sets with limited number of code samples.

An internal validity threat occurs if cause effect relationships between the dependent and independent variables in the experiments are not established properly. One limitation of this study that might affect reasoning about evidences more generally is that all code snippets in the generated data sets are implemented in Java language and some tactic implementations are based on common third party frameworks and libraries. Further research with more data sets is needed to repeat designed experiments for security controls implemented with other frameworks and programming languages.

\section{Conclusion}
\label{Conclusion}
This paper creates a comprehensive list of commonly used security controls and categorizes them into three abstract categories \ie \textit{Detect}, \textit{Prevent}, and \textit{React}. It pulls a set of tactic related and unrelated code snippets for these controls from the {\texttt{StackOverflow}} question and answer website to generate training and test data sets for conducted experiments. Using the Bidirectional Encoder Representations from Transformer (BERT) and the Tactic Detector (TD) from our prior work \cite{Mirakhorli2016} it shows that tactic related code snippets can be identified with quite high F-Measure values. 
In order to test the performance of the proposed approach on a real-life project, a case study is conducted where trained models are used to identify commonly used security tactics in the Liferay open source enterprise portal. Created data sets and scripts used to train the Tactic Detector and fine tune BERT are shared with the research community through GitHub, to make sure the results of the study are reproducible.


Given a set of code snippets, three experiments are designed to identify (1) the existence of an individual security tactic with a binary classification approach, (2) whether one of the controls in a given set is implemented or not, and (3) the type of implemented security tactic with a multi label classification approach. The results derived from the first two experiments show that fine-tuned BERT models and TD are able to achieve high Precision, Recall and F-Measure values while identifying whether security controls are implemented in a given code snippet or not. 
With BERT, the average F-Measure scores for the first two experiments are 0.97 and 0.96, respectively. Similarly, TD achieves average F-Measure values of 0.93 and 0.91, respectively. The results of the third experiment show that depending on the type of the security tactic and the size of its training set, the performance of the models may change. BERT and TD achieve F-Measure values up to 0.99 and 0.95, respectively using a multi-label classification approach. Finally, the case study conducted at the end shows the applicability of the developed approach on real-life projects.


\bibliographystyle{IEEEtran}
\bibliography{references.bib}

\end{document}